**High resolution Hall measurements across the VO$_2$ metal-insulator transition reveal impact of spatial phase separation**

**Tony Yamin[1,2], Yakov M. Strelniker[1] and Amos Sharoni[1,2].**

Department of Physics, Bar Ilan University, Ramat-Gan, Israel IL-5290002

Bar-Ilan Institute of Nanotechnology & Advanced Materials, Ramat-Gan, Israel IL-5290002

Many strongly correlated transition metal oxides exhibit a metal-insulator transition (MIT), the manipulation of which is essential for their application as active device elements. However, such manipulation is hindered by lack of microscopic understanding of mechanisms involved in these transitions. A prototypical example is VO$_2$, where previous studies indicated that the MIT resistance change correlate with changes in carrier density and mobility. We studied the MIT using Hall measurements with unprecedented resolution and accuracy, simultaneously with resistance measurements. Contrast to prior reports, we find that the MIT is not correlated with a change in mobility, but rather, is a macroscopic manifestation of the spatial phase separation which accompanies the MIT. Our results demonstrate that, surprisingly, properties of the nano-scale spatially-separated metallic and semiconducting domains actually retain their bulk properties. This study highlights the importance of taking into account local fluctuations and correlations when interpreting transport measurements in highly correlated systems.

Among the transition metal oxides - which are typically strongly correlated materials showing a plethora of phase transitions - many exhibit a metal-insulator transition (MIT)[1,2], the manipulation of which is an essential step towards using them in novel electronic devices, a field recently coined 'Mottronics'[3,4]. $VO_2$ is a prototypical MIT material, promising also for applications[5-7]. It has a critical temperature near room temperature, at 340K, where the resistance changes by over four orders of magnitude[2,8]. The MIT is accompanied by a structural phase transition from a low temperature monoclinic semiconductor (SC) to a metallic rutile and a hysteresis of a few degrees[2,9]. The transition in thin films and nano-rods has been shown to be spatially phase separated (SPS), where both SC and metallic 'puddles' coexist during the transition[10-14].

There is still much debate and contradicting evidence concerning the microscopic origin of the electronic and structural phase transitions and the mechanisms involved in these transitions, especially when considering different driving methods[15-20]. It is accepted that even if the transition is not purely electronic, Mott-Hubbard electron-electron correlations play an important role in the evolution of the phase transition and properties of the insulating state in $VO_2$[21-23]. This means that small changes to the carrier density (CD) can modify the electronic ground state and properties considerably[18,24,25]. Thus, accurate measurements of CD as a function of intrinsic changes or external stimuli can help to understand these systems, as exemplified in other correlated electron systems, such as cuprates[26], heavy fermions [27] interface of LAO/STO[28,29].

$VO_2$ shows a change in carrier density of ~4 orders between the low temperature SC and high temperature metallic states, responsible for the 4 orders resistivity change, with not much difference to mobility [30]. In spite of five decades of studies there are only scarce reports of Hall measurements of the CD in the close vicinity of the phase transition, especially for technologically relevant thin films of $VO_2$. These are lacking high temperature resolution and provide only a qualitative picture for the changes in CD and mobility near the phase transition, and usually disregard the intrinsic SPS[31,32]. A similar void exists in studies of CD in other MIT systems giving only qualitative pictures of changes in carrier density[33-35], and is probably due to experimental difficulties of attaining accurate Hall measurements during large resistivity changes over a small temperature range. The main problems are the need to stabilize temperature for each measurement to decrease noise, while avoiding overshoot of the set temperature which will introduce errors through hysteresis effects and need of large magnetic field sweeps to increase accuracy. All these make the measurement extremely time-consuming.

In this study we apply a simple modification to the Hall coefficient measurement scheme, which enables to increase temperature resolution and maintain high accuracy. We measure the Hall coefficient on single-phase thin $VO_2$ films, simultaneously with the resistance, across the temperature driven MIT. We find that, in contrary to previous reports, the resistance behaviour does not correspond to the changes in the Hall coefficient. We analyse the results in the context of spatial inhomogeneity of metallic and semiconducting domains during the transition, by introducing the exact relation theorem and effective medium approximation[36-38]. The hypothesis we use and confirm with the experimental results, is that even the small regions of spatially separated phases behave according to their bulk properties. Only at close vicinity to the percolation threshold we do observe a difference between the experimental results and the exact relation predictions, and discuss possible origins.

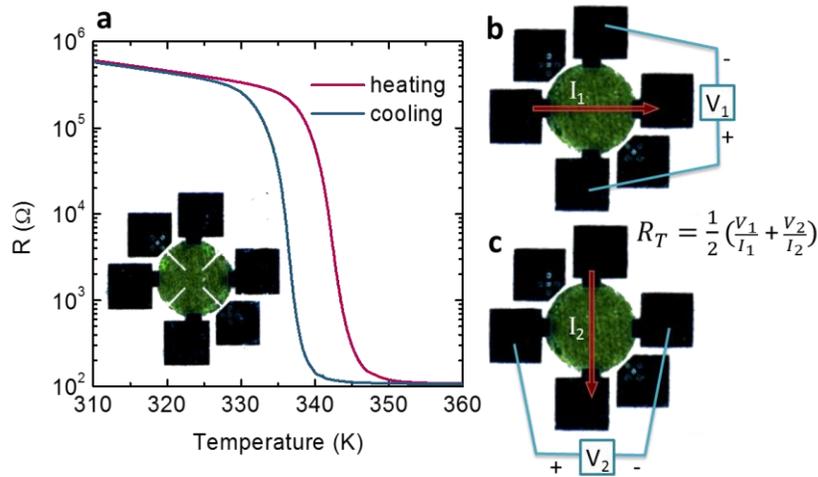

**Figure 1 | Device geometry and R vs. T data.** (**a**) Main: Resistance (log scale) vs. temperature measurement of the MIT in 65nm $VO_2$. Insets: final devices with 4 contacts. (**b**) and (**c**) The two configurations used to extract the offset-free transverse voltage measurement in the equation.

## Results

**High resolution Hall measurements.** Thin $VO_2$ films, 65nm thick, were deposited on R-cut sapphire substrates by reactive RF magnetron sputtering. The films show an MIT of over 3 decades (see Fig. 1) and were characterized as epitaxial, single phase with RMS roughness of 5nm (see supplementary Fig. 1). The films were patterned via reactive ion etching (RIE) into one of two commonly recommended geometries for Hall measurements[39]: a disc with diameter of 900μm, or a 4-leaf clover with identical outer diameter and inner spacing of 290μm. Optical images of the devices with contacts and bonding pads are presented in inset of Fig. 1a (clover) and Fig. 1b (disc). The Hall measurement for the disk and clover leaf geometries showed similar results, but the clover geometry, which is considered better is surprisingly noisier, possibly due to edge effects from the RIE process which are enhanced in this geometry (see supplementary Fig. 2). Herein, we present results of the disk geometry.

To avoid the large changing offsets in transverse voltage measurements[31], we apply the frequently used 'Onsager reciprocity method'[40,41], which is not common in studies of MIT materials. Here, for each field the Hall voltage is measured in two complementary configuration, exemplified in Fig. 1a and 1b, and using the reciprocity relation $R_{xy}(B) = R_{yx}(-B)$ (where $\overleftrightarrow{R}$ is the 2D resistivity matrix) an offset free transverse resistance, $R_T$, is extracted (see supplementary method and supplementary Fig. 3 for details)[42,43]. The Hall coefficient, $R_H$, is simply the slope of $R_T$ vs. magnetic field divided by the sample thickness. Strictly speaking, this method should work only for homogeneous materials, which is not the case in $VO_2$ because of the SPS. We measured the transverse resistance as a function of magnetic field for different temperatures across the MIT, presented in Fig. 2 for two temperatures, 330K and 340K in the right and left insets, accordingly. We draw attention that the offset is nearly zero. This means our device is large enough so the macroscopic measurements of the two configurations are on average the same. The carrier density we extract at 300K is $n = 7.9 \cdot 10^{18} cm^{-3}$, increasing to $n = 1 \cdot 10^{23} cm^{-3}$ above the MIT (at 350K); an increase of over 4 orders of magnitude. The results are in agreement with previous reports [31] (see also supplementary Fig. 5).

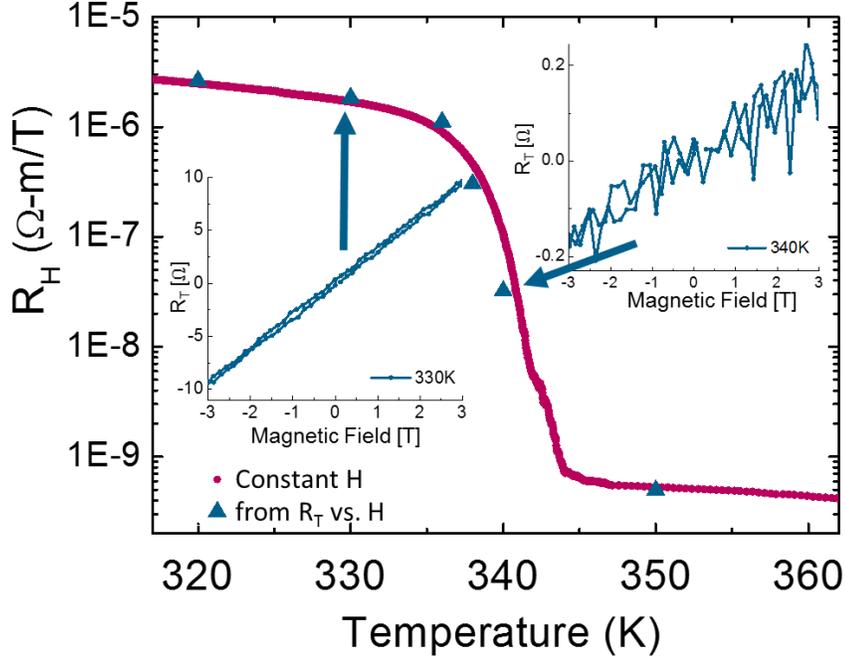

**Figure 2 | Hall measurements vs. field and temperature.** Main: $R_H$ vs. temperature for heating branch. Red circles - continuous measurement at constant magnetic field of 8T. Blue triangles - $R_H$ from magnetic field sweeps at constant temperature. Insets: $R_T$ vs. H for 330K (left) and 340K (right); arrows mark corresponding points in main panel.

Since we measure $R_T$ with nearly no offset, we can extract $R_H$ from a single high magnetic field, while ramping the temperature slowly and continuously. This enables us to perform high temperature resolution measurements of $R_H$ simultaneously with the longitudinal resistance, $R_{xx}$, across the MIT, overcoming the stabilization times and hysteresis problem. In Fig. 2 (main panel) we exemplify that the results for $R_H$ extracted from the continuous measurement (red circles) match the values from $R_T$ vs. H curves attained after stabilizing the temperature (blue triangles). Note this is a symbol plot, exemplifying the high temperature resolution.

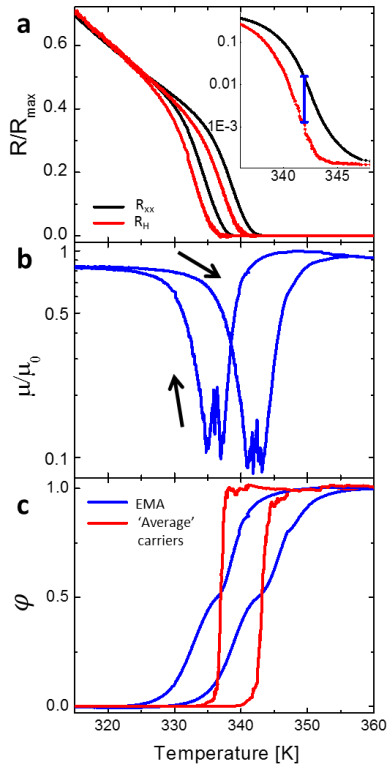

**Figure 3 | Bulk transport and Hall measurements.** (**a**) Simultaneous measurements of $R_{xx}$ (black curve) and $R_H$ (red curve) as a function of temperature, normalized to values at 310K. Inset is a zoom in, showing region where $R_H$ is an order of magnitude smaller than the longitudinal resistance. (**b**) Average Hall mobility vs. temperature. (**c**) Metallic fraction vs. temperature: Red curve- fraction extracted from CD. Blue curve- fraction extracted from resistance data and EMA.

Figure 3(a) shows a simultaneous measurement of $R_{xx}$ (black curve) and $R_H$ (red curve) vs. temperature, each normalized to its value measured at 310K. The main panel is in linear scale and the inset in log scale and smaller temperature range. At low temperatures, when the sample is semiconducting, the two curves are identical, indicating that resistance change is due to increase in carrier concentration and no evident change in mobility (since $R_{xx} \propto 1/n$ and $n \propto 1/R_H$). Also after the MIT, when the sample is metallic, the measurements coincide. But, during the MIT and for a wide temperature range there is a large difference between the two, of over 1 order of magnitude, as indicated by the blue bar in the inset. If one does not have any additional information on the material measured, the first explanation will likely be that the phase transition is accompanied by a large change in mobility, as suggested previously[32]. In Fig. 3b we plot the relative change in mobility as a function of temperature required to reconcile the Hall measurement. Since the mobility in both the SC and the metallic pure phases are of similar value, the behaviour cannot be explained in terms of an averaged mobility. The results could also be interpreted as a result of phase boundaries of the SPS during the MIT[10-12]. Other conclusions could be, for example, that the mobility decrease comes from enhanced scattering from SC-metal interfaces, which would place the minimum mobility where the most separation occurs[32]. Below we show the results are actually a direct outcome of the SPS transition with no need to assume a mobility decrease.

**Analysis in the framework of the exact relation theorem.** First, we exemplify that the Hall measurement does not relate to the mobility-weighted average carrier density of the coexisting carrier types of SC and metallic phases. For this we extract in two ways the volume fraction of the metallic phase ($\varphi$) across the MIT. Once- directly from the Hall coefficient. On average the carrier density is $n_{avg} = \varphi n_m + (1-\varphi)n_{SC}$, $n_M$ is the constant metallic CD measured at high temperatures and $n_{SC}$ is the temperature dependant CD of the SC extrapolated from the low temperature behaviour. We find $\varphi$ that fulfils $R_H = -\frac{1}{e[(1-\varphi)n_{SC}+\varphi n_m]}$. Second, $\varphi$ is extracted from the longitudinal resistance using the 2D effective medium approximation (EMA). One can consider the spatially separated domains as forming a percolation network, shown to provide a good estimation of the spatially separated phase fraction in $VO_2$ during the MIT[44,45]. In the low magnetic field approximation, which is the case here, the 2D EMA matches the EMA in absence of magnetic field[36,37]: $(\sigma_m - \sigma_e)/(\sigma_m + \sigma_e) + (1 - \varphi)(\sigma_{SC} - \sigma_e)/(\sigma_{SC} + \sigma_e) = 0$, from which we can find $\varphi$ again. Here $\sigma_{SC,m,e}$ are the conductivities of the SC phase, metallic phase, and measured effective longitudinal conductivity. We can use $\sigma_{SC,m,e} = C \cdot 1/R_{SC,m,xx}$ and since C is a constant geometrical factor it is cancelled out. Resistances of the metallic and SC phases are extracted in a similar fashion as done for carrier densities above. Figure 3c is a plot of the comparison, showing a large discrepancy between the metallic fraction extracted from the CD, in red, and the EMA, in blue. Therefore, the Hall measurement cannot be interpreted as an averaged charge carrier density.

We now analyse the results using the 2D exact relation theorem. It determines that in a two component composite there exists an exact relation between the effective values of the 2D conductivity tensor components and conductivity tensor components of each of the composite constituents[38,46-48]. We assume that the SC and metallic spatially separated phases have different conductivities and different Hall coefficients. We emphasize that the hall coefficients for the SC ($R_{H,SC}$) and metallic phases ($R_{H,m}$) are not fitting parameters. They are extracted from the measured single phase regime, i.e. low temperatures for SC and high for metallic, see supplementary Fig. 6. The expected Hall coefficient according to the exact relations fulfils (for low magnetic field approximation):

$$(\sigma_{H,e} - \sigma_{H,SC})/(\sigma_{H,m} - \sigma_{H,SC}) = (\sigma_e^2 - \sigma_{SC}^2)/(\sigma_m^2 - \sigma_{SC}^2)$$

Here $\sigma_{H,i}$ ($i = SC, m, e$) are the Hall conductivities of the SC phase, metallic phase, and effective Hall conductivity, i.e. expected from a measurement. We can use the relation: $\sigma_{H,i} = R_H \sigma_{XX}^2 B$ since $R_H B \ll R_{XX}$.

Figure 4 shows the results of the exact relation in terms of the Hall coefficient (blue curves) compared to the experimental data (red curves). Figure 4a is a semi log plot, and Fig. 4b and 4c are linear plots focusing on different regions. As evidenced in Fig. 4a and 4b there is very good agreement between the measured $R_H$ and $R_H$ extracted from the exact relations. There is only one region where the exact relation shows a deviation from the experimental data, magnified in Fig. 4c, where the experimental $R_H$ is ~2 time larger, not much relative to the 4 order of change the exact relation captures successfully. We find that, interestingly, this area corresponds to the percolation threshold, i.e. the point where long-range connectivity occurs for the metallic puddles in the spatially-separated domains network. Using the 2D EMA, we find the temperature of the samples percolation threshold which occurs at a phase fraction $\varphi = 0.5$ [37,49], marked in Fig. 4c by the arrow. There are a few possible reasons for this difference, for example: (1) The exact relation theorem is not accurate near the percolation threshold. In this case, it is not clear why it results in under-estimation. (2) The assumption that the bulk properties of the SC phase hold along the MIT is not correct near percolation, when the puddles can be of nanometre size. (3) The Onsager method requires that the sample can be considered homogenous on a relevant small length-scale, but near percolation threshold there is a divergence. Further study and theoretical predictions are needed in order to resolve this issue, which is beyond the scope of this article.

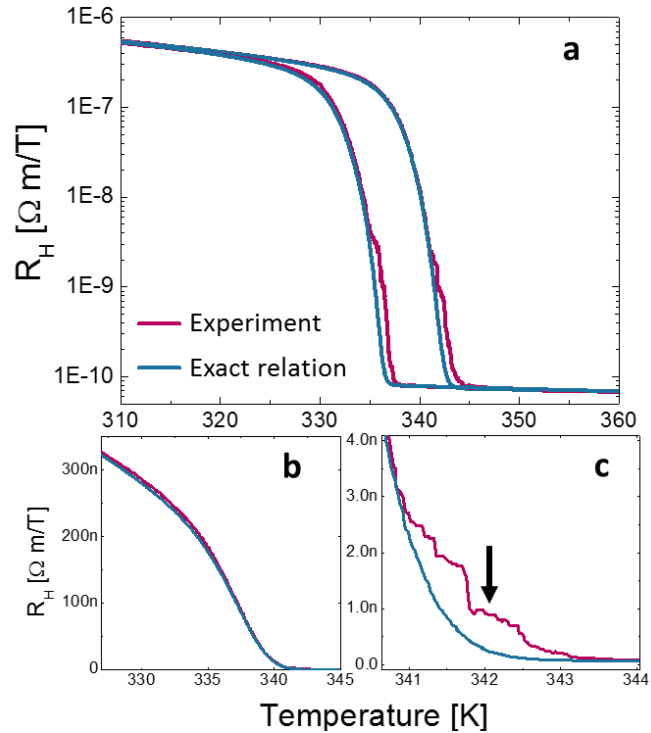

**Figure 4 | Hall coefficient vs. temperature**. Red curve is experimental, and blue- via exact relation theorem. (**a**) Semi log plot of entire range. (**b**) Linear scale showing excellent fit for large Hall coefficient. (**c**) Zoom on the only region where there is discrepancies. Arrow indicates percolation threshold.

## Discussion

We recall that the only postulate used to achieve the quantitative agreement of the exact relation theorem is that throughout the spatial phase separation both the metallic and SC puddles have the same properties of the fully developed phase. This indicates that during almost the entire MIT (excluding the percolation threshold), even for rather small SC puddles (few tens of nanometres[50]) they behave with the same temperature dependence of the fully SC film, and similarly for the metallic portion. In addition, this signifies that carrier mobility is almost constant along the phase transition, with no noticeable phase-boundary scattering, in correspondence with mobility of SC and metallic phases being almost the same. These results highlight the importance of considering the effects of inhomogeneity (in space or time) when analysing properties of highly correlated transition metal oxides. Inhomogeneity is emerging as an intrinsic properties in many such systems, may have considerable shift in interpretation of results, i.e. resolving the mobility change presented here. There are additional examples for the importance of considering inhomogeneity, such as role of local moments in magnetic properties of LAO/STO heterostructures [28], resolving dielectric properties of tunnel barriers [51] and analysing properties of disordered superconductors [52]. Thus in correlated systems, the presence of inhomogeneity cannot be assumed to average out when interpreting macroscopic measurements, since small fluctuations can cause large changes of the measured properties in these materials.

On a different note, the $VO_2$ system is found to be appealing for the study of Hall and other magneto-transport properties in percolation systems. Mainly since temperature control enables high control of fraction parameter, and also because the $VO_2$ regular magnetoresistance is very small, 0.09% at 12T[31]. Most experimental studies of the exact relation or EMA were done on a limited number of individually prepared samples [53,54]. Here we showed that the exact relation is excellent in capturing almost the entire fraction range. Research can be naturally expanded to study other aspects, including finite size systems, crossover from 2D to 3D and critical exponents, to name a few.

In summary, we preformed high resolution Hall measurements of $VO_2$ thin films and followed the evolution of Hall coefficient and sample resistance across the MIT simultaneously and accurately. We show, using the exact relation theorem, that the Hall coefficient does not correspond to changes in the VO2 carrier density, but is rather a macroscopic manifestation of the spatial phase separation accompanying the MIT. Our results further indicate that the small SPS puddles maintain their bulk properties except, possibly, at the percolation threshold. Our study highlights the importance of spatial changes and local fluctuations when analysing macroscopic properties of MIT materials and other highly correlated transition metal oxides.

**Methods**

**Film deposition**. $VO_2$ films, 65nm thick, were deposited on R-cut (1-102) sapphire substrates by reactive RF magnetron sputtering in a high vacuum chamber (AJA Int.) with base pressure of $1 \times 10^{-8}$ Torr and Ar/O$_2$ mixture from a nominal V$_2$O$_3$ target. Optimal growth occurred at a substrate temperature of 600C, total pressure of 3mtorr, oxygen partial pressure of 2% and a growth rate of 0.17 Å/s. Under these conditions the films grew smoothly, epitaxial and single-phased, as confirmed by AFM, X-ray diffraction and HR-SEM. More details concerning the preparation and characterization of these films can be found in the supplementary material and previous work [55].

**Device fabrication**. The films were patterned into the disk geometries for Hall measurements using a photolithography mask and reactive ion etching (RIE) (ICP-RIE, SPTS Inc.). Etching parameters were- pressure of 20mTorr composed of 67% SF$_6$ and 33% Ar resulting in an etch rate ~ 30nm/min.. An additional step of photolithography and lift-off was used to align vanadium contacts and large bonding pads, 120nm thick, deposited by sputtering.

**Magneto-electrical measurements**. All measurements were performed in a commercial cryostat (PPMS, Quantum Design). We measured resistance and transverse voltage using a Keithley 6211 high impedance current source and Keithley 2182A nanovoltmeter, while controlling the measurement configuration with a fast switch system, Keithley 7001 with 10x4 matrix card. Magnetic field dependent Hall measurements were done by first stabilizing the temperature (around 30 minutes) and then following the reciprocity method to measure R$_T$ depicted in Fig. 1b and 1c. Magnetic fields were swept between -3T and 3T. The temperature was ramped at rates between 1K/min to 0.1K/min (near the MIT), verifying the rate is low enough so that there is only a small resistance change between adjacent measurements, see supplementary methods and supplementary Fig.3. We performed slow temperature sweeps at constant magnetic fields of 8T, 0T and -8T to extract the Hall coefficient more accurately, shown in supplementary Fig. 4.

### References

1.  Imada, M., Fujimori, A. & Tokura, Y. Metal-insulator transitions. *Rev. Mod. Phys.* **70,** 1039 (1998).
2.  Morin, F. J. Oxides Which Show a Metal-to-Insulator Transition at the Neel Temperature. *Phys. Rev. Lett.* **3,** 34 (1959).

3. Inoue, I. H. & Rozenberg, M. J. Taming the Mott Transition for a Novel Mott Transistor. *Adv. Funct. Mater*. **18,** 2289-2292 (2008).
4. Hwang, H. Y. *et al.* Emergent phenomena at oxide interfaces. *Nat. Mater.* **11,** 103-113 (2012).
5. Shukla, N. *et al.* A steep-slope transistor based on abrupt electronic phase transition. *Nat. Commun.* **6,** 7812 (2015).
6. Zhou, Y. & Ramanathan, S. Correlated Electron Materials and Field Effect Transistors for Logic: A Review. *Crit. Rev. Solid State Mater. Sci.* **38,** 286-317 (2013).
7. Appavoo, K. *et al.* Ultrafast Phase Transition via Catastrophic Phonon Collapse Driven by Plasmonic Hot-Electron Injection. *Nano Lett.* **14,** 1127-1133 (2014).
8. Berglund, C. N. & Guggenheim, H. J. Electronic Properties of $VO_2$ near the Semiconductor-Metal Transition. *Phys. Rev.* **185,** 1022 (1969).
9. Goodenough, J. B. The two components of the crystallographic transition in $VO_2$. *J. Solid State Chem.* **3,** 490-500 (1971).
10. Sharoni, A., Ramirez, J. G. & Schuller, I. K. Multiple avalanches across the metal-insulator transition of vanadium oxide nanoscaled junctions. *Phys. Rev. Lett.* **101,** 0264041-0264044 (2008).
11. Seal, K. *et al.* Resolving transitions in the mesoscale domain configuration in $VO_2$ using laser speckle pattern analysis. *Sci. Rep.* **4,** 6259 (2014).
12. Qazilbash, M. M. *et al.* Mott Transition in $VO_2$ Revealed by Infrared Spectroscopy and Nano-Imaging. *Science* **318,** 1750-1753 (2007).
13. O'Callahan, B. T. *et al.* Inhomogeneity of the ultrafast insulator-to-metal transition dynamics of $VO_2$. *Nat. commun.* **6,** 6849 (2015).
14. Yee, C.-H. & Balents, L. Phase Separation in Doped Mott Insulators. *Phys. Rev. X* **5,** 021007 (2015).
15. Zimmers, A. *et al.* Role of Thermal Heating on the Voltage Induced Insulator-Metal Transition in $VO_2$ *Phys. Rev. Lett.* **110,** 056601 (2013).
16. Xue, X. *et al.* Photoinduced insulator-metal phase transition and the metallic phase propagation in $VO_2$ films investigated by time-resolved terahertz spectroscopy. *J. Appl. Phys.* **114,** 193506 (2013).
17. Kim, H.-T. *et al.* Monoclinic and Correlated Metal Phase in $VO_2$ as Evidence of the Mott Transition: Coherent Phonon Analysis. *Phys. Rev. Lett.* **97,** 266401-266404 (2006).
18. Chen, F. H. *et al.* Control of the Metal–Insulator Transition in $VO_2$ Epitaxial Film by Modifying Carrier Density. *ACS Applied Materials & Interfaces* **7,** 6875-6881 (2015).
19. Brockman, J. S. *et al.* Subnanosecond incubation times for electric-field-induced metallization of a correlated electron oxide. *Nat Nano* **9,** 453-458 (2014).
20. Liu, M. *et al.* Phase transition in bulk single crystals and thin films of $VO_2$ by nanoscale infrared spectroscopy and imaging. *Phys. Rev. B* **91,** 245155 (2015).
21. Zylbersztejn, A. & Mott, N. F. Metal-insulator transition in vanadium dioxide. *Phys. Rev. B* **11,** 4383–4395 (1975).
22. Biermann, S., Poteryaev, A., Lichtenstein, A. I. & Georges, A. Dynamical Singlets and Correlation-Assisted Peierls Transition in $VO_2$. *Phys. Rev. Lett.* **94,** 026404 (2005).
23. Aetukuri, N. B. *et al.* Control of the metal-insulator transition in vanadium dioxide by modifying orbital occupancy. *Nat. Phys*. **9,** 661-666 (2013).
24. Jeong, J. *et al.* Suppression of Metal-Insulator Transition in $VO_2$ by Electric Field–Induced Oxygen Vacancy Formation. *Science* **339,** 1402-1405 (2013).
25. Esfahani, D. N., Covaci, L. & Peeters, F. M. Field effect on surface states in a doped Mott-insulator thin film. *Phys. Rev. B* **87,** 035131 (2013).

26. Takagi, H., Uchida, S. & Tokura, Y. Superconductivity produced by electron doping in $CuO_2$ layered compounds. *Phys. Rev. Lett.* **62,** 1197-1200 (1989).
27. Paschen, S. *et al.* Hall-effect evolution across a heavy-fermion quantum critical point. *Nature* **432,** 881-885 (2004).
28. Kalisky, B. *et al.* Critical thickness for ferromagnetism in $LaAlO_3$/$SrTiO_3$ heterostructures. *Nat. Commun.* **3,** 922 (2012).
29. McGuire, J. A. & Shen, Y. R. Tunable Quasi–Two-Dimensional Electron Gases in Oxide Heterostructures. *Science* **313,** 1942-1945 (2006).
30. Rosevear, W. H. & Paul, W. Hall Effect in $VO_2$ near the Semiconductor-to-Metal Transition. *Phys. Rev. B* **7,** (1973).
31. Ruzmetov, D., Heiman, D., Claflin, B. B., Narayanamurti, V. & Ramanathan, S. Hall carrier density and magnetoresistance measurements in thin-film vanadium dioxide across the metal-insulator transition. *Phys. Rev. B* **79,** 153107 (2009).
32. Fu, D. *et al.* Comprehensive study of the metal-insulator transition in pulsed laser deposited epitaxial $VO_2$ thin films. *J. Appl. Phys.* **113,** 043707 (2013).
33. Saito, Y. & Iwasa, Y. Ambipolar Insulator-to-Metal Transition in Black Phosphorus by Ionic-Liquid Gating. *ACS Nano* **9,** 3192-3198 (2015).
34. Hauser, A. J. *et al.* Correlation between stoichiometry, strain, and metal-insulator transitions of $NdNiO_3$ films. *Appl. Phys. Lett.* **106,** 092104 (2015).
35. Zhang, L., Gardner, H. J., Chen, X. G., Singh, V. R. & Hong, X. Strain induced modulation of the correlated transport in epitaxial $Sm_{0.5}Nd_{0.5}NiO_3$ thin films. *J. Phys. Condens. Matter.* **27,** 132201 (2015).
36. Bergman, D. J. & Strelniker, Y. M. Magnetotransport in conducting composite films with a disordered columnar microstructure and an in-plane magnetic field. *Phys. Rev. B* **60,** 13016 (1999).
37. Landauer, R. The Electrical Resistance of Binary Metallic Mixtures. *J. Appl. Phys.* **23,** 779 (1952).
38. Shklovskii, B. I. Critical behavior of the Hall coefficient near the percolation threshold. *Sov. Phys. JETP* **45,** 152 (1977).
39. van der Pauw, L. J. A method of measuring specific resistivity and hall effect of discs of arbitray shape. *Philips Res. Rep.* **13,** 1-9 (1958).
40. Cornils, M. & Paul, O. Reverse-magnetic-field reciprocity in conductive samples with extended contacts. *J. Appl. Phys.* **104,** 02451-02457 (2008).
41. Cornils, M., Rottmann, A. & Paul, O. How to Extract the Sheet Resistance and Hall Mobility From Arbitrarily Shaped Planar Four-Terminal Devices With Extended Contacts. *IEEE Trans. Electron Dev.* **57,** 2087-2097 (2010).
42. Onsager, L. Reciprocal Relations in Irreversible Processes. II. *Phys. Rev.* **38,** 2265-2279 (1931).
43. Casimir, H. On Onsager's Principle of Microscopic Reversibility. *Rev. Mod. Phys.* **17,** 343-350 (1945).
44. Hood, P. J. & DeNatale, J. F. Millimeter-wave dielectric properties of epitaxial vanadium dioxide thin films. *J. Appl. Phys.* **70,** 376 (1991).
45. Rozen, J., Lopez, R., Haglund, J. R. F. & Feldman, L. C. Two-dimensional current percolation in nanocrystalline vanadium dioxide films. *Appl. Phys. Lett.* **88,** 081902-081903 (2006).
46. Dykhne, A. M. Conductivity of a two-dimensional two-phase system. *Sov. Phys. JETP* **32,** 63 (1971).
47. Bergman, D., Duering, E. & Murat, M. Discrete network models for the low-field Hall effect near a percolation threshold: Theory and simulations. *J. Stat. Phys.* **58,** 1-43 (1990).
48. Bergamn, D. J. & Stroud, D. Physical properties of macroscopically inhomogeneous media. *Solid State Phys.* **46,** 147-269 (1992).

49. Bergman, D. J. & Strelniker, Y. M. Magnetotransport in conducting composite films with a disordered columnar microstructure and an in-plane magnetic field. *Phys. Rev. B* **60,** 13016-13027 (1999).
50. Frenzel, A. *et al.* Inhomogeneous electronic state near the insulator-to-metal transition in the correlated oxide $VO_2$ *Phys. Rev. B* **80,** 115115 (2009).
51. Miller, C. W., Li, Z.-P., Åkerman, J. & Schuller, I. K. Impact of interfacial roughness on tunneling conductance and extracted barrier parameters. *Appl. Phys. Lett.* **90,** 043513 (2007).
52. Dubi, Y., Meir, Y. & Avishai, Y. Nature of the superconductor–insulator transition in disordered superconductors. *Nature* **449,** 876-880 (2007).
53. Dai, U., Palevski, A. & Deutscher, G. Hall effect in a three-dimensional percolation system. *Phys. Rev. B* **36,** (1987).
54. Zhang, X. X. *et al.* Giant Hall Effect in Nonmagnetic Granular Metal Films. *Phys. Rev. Lett.* **86,** 5562-5565 (2001).
55. Yamin, T., Havdala, T. & Sharoni, A. Patterning of epitaxial $VO_2$ microstructures by a high-temperature lift-off process. *Mater. Res. Express* **1,** 046302 (2014).

**Acknowledgments**

We thank A. Yoffe and the Nano-fabrication unit, Chemical Research Support Department WIS, for help with the RIE process. This research was supported by the ISRAEL SCIENCE FOUNDATION (grant No. 727/11).

**Author Contributions**

T.Y. grew the vanadium oxide films, fabricated the Hall devices and carried out the electrical measurements. T.Y. and Y.M.S performed the numerical simulations. A.S. and T.Y. conceived the experiment and wrote the manuscript together. All authors discussed the results and commented on the manuscript.

**Additional information**

**Supplementary Information** accompanies this paper

**Competing financial interests:** The authors declare no competing financial interests.